# Towards Automatically Pruning Logging Code with Coding Agents: How Far Are We?

He Yang Yuan
York University
Toronto, Canada
yuanh@yorku.ca

Haonan Zhang
University of Waterloo
Waterloo, Canada
haonan.zhang@uwaterloo.ca

Xin Wang
HKUST (Guangzhou)
Guangzhou, China
xwang496@connect.hkust-gz.edu.cn

An Ran Chen
University of Alberta
Edmonton, Canada
anran6@ualberta.ca

Kisub Kim
DGIST
Daegu, Korea
kisub.kim@dgist.ac.kr

Zishuo Ding
HKUST (Guangzhou)
Guangzhou, China
zishuoding@hkust-gz.edu.cn

Zhenhao Li
York University
Toronto, Canada
lzhenhao@yorku.ca

## Abstract

Logging code supports debugging, monitoring, and software maintenance, and developers continually revise it as software evolves. Excessive logging code can add noise to logs, impose runtime overhead, and obscure information needed for diagnosis. While prior research has extensively studied logging code generation and modification, how developers remove logging code remains comparatively underexplored. In this paper, we conduct a comprehensive study of developer logging removal practices and explore the potential of using coding agents to automate this task. We extract and manually validate logging removal cases from Python and Java repositories and derive 10 removal patterns and 11 removal reasons that characterize how and why logging code was removed in real-world software changes. We further construct **LogRem**, a dataset of 387 real-world cases covering direct logging statement removal, logging infrastructure removal, and logging replacement. We evaluate four coding agents with multiple model settings and compare their outputs with real-world changes that were accepted and merged into the corresponding projects. Although 95.6% to 100.0% of outputs pass validity checks, only 11.1% to 19.6% remove the same logging code as the corresponding real-world change while preserving unrelated code. Agents differ from real-world changes through missed removals, extra removals, and unrelated code edits, with substantial variation across logging removal categories and trajectories. Execution cost also varies widely, but higher cost does not consistently yield closer alignment with real-world changes. In a further comparison of two different forms of input context, commit messages and developer discussions lead to the largest improvements in alignment with real-world changes. Guidance based on our taxonomy of logging removal patterns and reasons consistently reduces runtime. Overall, our study establishes logging removal as a distinct software maintenance task and shows that reliable automation depends on accurately determining removal scope while preserving necessary code. To the best of our knowledge, this is the first study to examine logging code removal from this perspective.

## 1 Introduction

Software systems rely on logging to make their runtime behavior observable [65]. Logging statements record execution events, program state, and failure information [17, 19] that developers use for debugging [27], fault diagnosis [1, 70], monitoring [6, 55], testing [68], and other maintenance activities [22, 38]. However, useful logging requires balancing observability against the cost of producing and processing log data. Excessive logging can increase runtime overhead, generate redundant information, and make important diagnostic information harder to identify [14, 28].

These concerns are recognized in both industrial practice and academic research. AWS recommends limiting excessive logging because it can affect performance, increase storage and processing costs, and make important events harder to detect [54]. Microsoft recommends controlling logging verbosity in production and warns that logging operations can incur computation even when the corresponding log level is disabled [45, 46]. Google similarly recommends reducing unnecessary log volume to control ingestion, storage, and downstream processing costs [20]. Academic studies identify pre-logging overhead [71] and unnecessary exception stack trace logging [31] as practical logging quality problems.

As software evolves, the value of individual logging statements can also change. Developers revise logging while fixing bugs, changing features, refactoring implementations, and redesigning error handling [4, 40, 67]. Logging that was useful when introduced may later become redundant, costly, or inconsistent with the current implementation. Maintaining logging code therefore requires deciding both what information should continue to be recorded and what logging should be removed.

Figure 1 shows a real logging removal example from an open source project **vLLM** [59]. The project provides a *"--disable-log-requests"* option, but developers observed that requests were still logged when this option was enabled. After code inspection, the developer identified two hard coded logging statements in the OpenAI compatible API endpoints that bypassed the configuration and recorded the complete incoming requests. The developer removed these two logging statements while preserving the surrounding code. This example illustrates that logging removal can require developers to trace unexpected logging behavior to specific source locations and determine a precise removal scope without affecting other logging or program behavior.

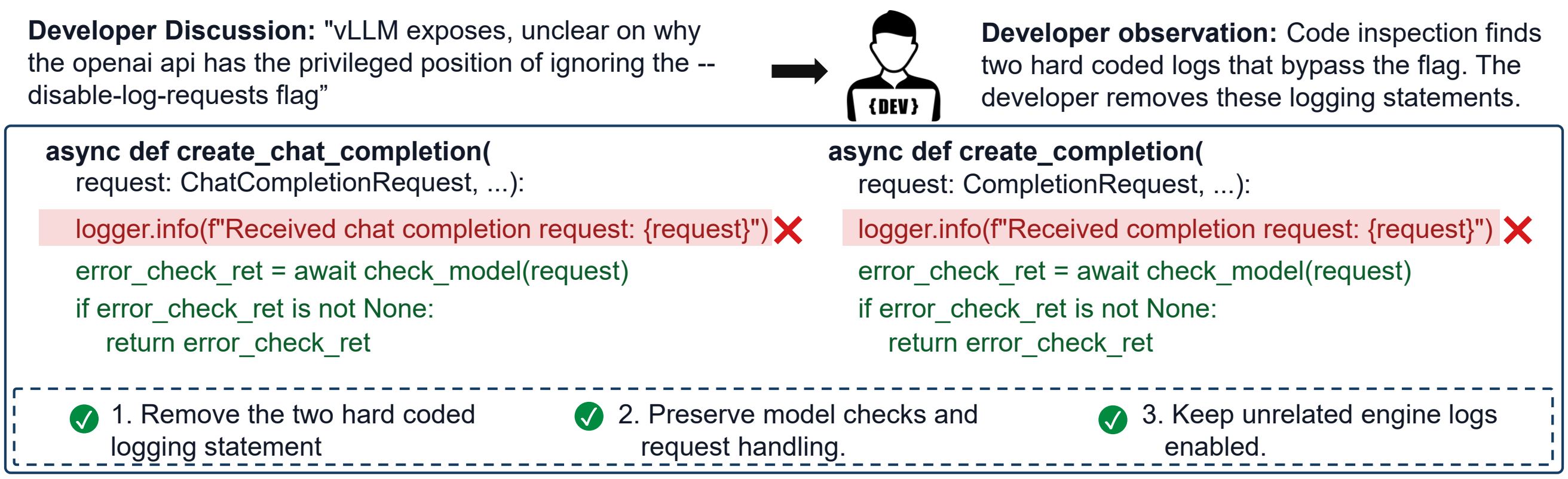

**Figure 1: A logging removal change that addresses requests being logged despite the *--disable-log-requests* option. The developer removes two statements that bypass the option while retaining model checks, request handling, and unrelated engine logs.**

Prior research has extensively studied logging evolution and automated support for logging location, level, content, generation, and maintenance [26, 30, 32–34, 39, 41, 43, 63, 73]. Recent work has also explored coding agents for generating and maintaining logging statements [50, 72]. In comparison, logging removal remains underexplored. Although deletion appears in studies of logging evolution [30], existing work provides limited understanding of why developers remove logging code, what related code changes are involved, and how such changes could be automated.

Logging removal also requires more than deleting an individual logging statement. Developers may need to determine which logging code should be removed while preserving other logging and surrounding code. The change may also require cleaning up logging infrastructure, or replacing logging with another reporting or error-handling mechanism. Recent coding agents can address repository-level maintenance tasks such as issue resolution, code editing, and repair [9, 42, 60, 64], but it remains unclear whether they can make these logging removal decisions when the expected removal scope is not specified in advance. This motivates us to examine how developers perform logging removal and how closely coding agents can reproduce such changes.

In this work, we study how and why logging code was removed and examine how coding agents perform on these changes. We mine 14,931 file-level logging changes from 240 Python and Java repositories and identify 1,564 removals, which account for 10.5% of the collected changes, indicating that logging removal is prevalent. We manually inspect these cases and validate 735 logging removals, the remaining changes are additions, modifications, relocations and changes that do not pass manual validation. Our analysis shows that logging removal follows several patterns and occurs for a range of removal reasons. Developers may remove individual logging statements, clean up logging infrastructure, replace logging with another behavior, or remove logging together with larger parts of the implementation. We organize these cases into 10 removal patterns and 11 removal reasons. To evaluate coding agents on cases where logging removal can be studied separately from broader code removal, we select four patterns that involve direct logging statement removal, logging infrastructure removal, or replacement of logging with another behavior. This results in 387 cases, including 279 direct logging statement removals, 49 logging infrastructure removals, and 59 logging replacements. We construct the dataset **LogRem** using these cases. For each case, we retain the original real-world change and compare the agent output with it to determine whether the agent removes the same logging code and leaves unrelated code unchanged.

Our results show that coding agents usually return valid outputs, but their logging removal decisions often differ from those made in the corresponding real-world change. When given only the task description and source code, only 11.1% to 19.6% of outputs remove the same logging code as corresponding real-world changes while preserving unrelated code. Performance also differs substantially across logging removal categories: direct logging statement removal reaches 19.4% alignment, while logging replacement reaches only 5.9%. Execution cost varies widely across configurations and does not consistently correspond to closer alignment with real-world changes. Additional context also changes agent performance. Commit messages and developer discussions provide the largest alignment gains, increasing alignment with real-world changes by approximately 9-11%. Guidance based on our taxonomy of logging removal patterns and reasons consistently reduces runtime.

Overall, our study establishes logging removal as a distinct maintenance problem centered on inferring the intended change scope while preserving unrelated code. To the best of our knowledge, this is the first study to examine logging code removal from this perspective. Our findings show that producing valid code is easier than making precise removal decisions. These results can inform future work on more scope-aware and efficient automation of logging maintenance. We summarize the contributions of this paper as follows:

- We conduct a comprehensive study of logging removal as a distinct software maintenance activity. Based on manually validated real-world removals, we characterize how and why logging code was removed in real-world changes through 10 removal patterns and 11 removal reasons.

- We construct **LogRem** by turning 387 real-world logging removal cases into tasks for coding agents. The tasks cover direct logging statement removal, logging infrastructure cleanup, and logging replacement, and each includes the source code, a task description, and the corresponding real-world change for reference.
- Our experimental results show that coding agents face different challenges across logging removal tasks. They align more closely with real-world changes when directly removing logging statements, but struggle more when logging must be replaced. Our analysis of agent actions also reveals when they miss intended removals or remove additional code, which can help identify where future approaches need to improve.

# 2 Background and Related Work

## 2.1 Logging Practices and Evolution

Empirical studies have extensively examined how developers use and maintain logging code. Prior work shows that logging statements evolve frequently in open source systems and mobile applications, and has investigated which characteristics are associated with later logging changes [4, 26, 65, 67]. Other studies focus on quality problems in existing logging code, including logging related issues introduced during maintenance [5, 38, 40, 61, 66]. Researchers have also studied the factors that influence developer logging decisions, such as diagnostic benefits, maintenance cost, and developer experience [29, 53], as well as logging practices in specific settings including test code, machine learning applications, and Java logging utilities [6, 18, 68]. More broadly, Gu et al. [22] organize prior research around why, where, what, and how well to log, and reports substantially more attention to logging placement and content than to the motivations behind logging decisions. Prior work on logging evolution also includes deletion as one type of logging change [30], but removal has generally been studied as part of broader logging evolution rather than as a maintenance problem of its own. However, logging removal nevertheless appears repeatedly in software evolution histories. This motivates our focus on logging removal as a distinct maintenance problem. Our work focuses specifically on how logging code is removed and why these changes were made in real-world software projects.

## 2.2 Automated Logging Generation and Maintenance

Research on automated logging has addressed several decisions involved in creating and maintaining logging statements. Learning based techniques have been proposed for recommending logging locations and levels, while other approaches generate logging messages or complete logging statements from source code [15, 16, 37, 39, 41, 43, 63]. Software history has also been used to learn recurring logging revisions and support the maintenance of existing statements [32, 33]. More recently, LLM based techniques have been applied to logging generation and maintenance by incorporating program context, identifying logging gaps, and repairing defects in existing logging statements [34–36, 73]. Evaluation has progressed beyond textual similarity as well. For example, Tan et al. [57] evaluate generated logging statements using compilation and runtime behavior. However, existing techniques primarily focus on adding logging code, generating its content, or modifying existing statements. Logging removal remains less studied, particularly when removal requires related cleanup or replacement behavior. Our work addresses this gap by studying coding agents on logging removal and comparing their outputs with observed real-world changes.

## 2.3 Coding Agents and Evaluations

LLM-based coding agents perform software maintenance by interacting with repositories, inspecting source files, editing code, running commands, and using execution feedback to revise their solutions [25, 60, 64]. Recent studies have also moved beyond final patch correctness to examine agent trajectories, repository exploration, and tool use during task execution [3, 8, 23, 69]. Specialized evaluations further study agent behavior for particular software engineering concerns, such as security [7]. Logging has recently become a target for coding agents. Zhong et al. [72] use multiple agents to determine whether logging is needed and generate logging statements, while Ouatiti et al. [50] study how coding agents create and maintain logging code in real pull-requests. Our study explores how coding agents perform on logging removal and examines both their final outputs and their trajectories.

# 3 Characterizing Real-World Logging Removals

Figure 2 summarizes the overall workflow of our study. In this section, we investigate how and why logging code was removed in real-world software changes. As shown in Stage 1 of Figure 2, we mine the histories of Python and Java repositories, identify logging related changes, and manually validate removal cases. We then analyze these cases to derive removal patterns and removal reasons.

## 3.1 Repository Mining and Logging Change Identification

We mine Python and Java repositories from GitHub. Following prior studies on repository selection [11, 12, 51], we exclude forks, archived repositories, templates, repositories without a license, and repositories that cannot be checked out at the target commits. We also require production source files in the target language, and more than five contributors. After applying these criteria, we rank the remaining repositories by star count and select the top 200 repositories for each language, resulting in an initial collection of 400 repositories. For each repository, we collect the commits between June 30, 2021 and June 30, 2026. Following prior work [62], we exclude merge commits and inspect changed Python and Java files for logging modifications. For each case, we retain the repository and commit identifiers, file path, commit metadata, pull-request information when available, the complete source code before and after the commit, and the unified diff.

We identify logging changes by combining static source analysis for each language with the unified diff. For Python, we use AST analysis to identify calls to common logging receivers, including "`logging`", "`logger`", "`self.logger`", "`_logger`", and "`_log`", as well as logging configuration and logger management operations. For Java, we exclude text inside comments and string literals, then identify logging operations from commonly used frameworks, including Java Util Logging, SLF4J, Log4j, Logback, Android Log, and Timber. For both languages, we consider a logging occurrence only when it overlaps a line added, deleted, or modified in the unified diff. After removing incomplete and duplicate entries, we obtain 14,931 file level changes from 240 repositories. We classify these changes based on how they affect logging, resulting in 3,427 additions, 1,564 removals, 9,556 modifications, and 384 relocations. We then examine the 1,564 removal cases in detail to verify the changes and characterize their removal patterns and reasons.

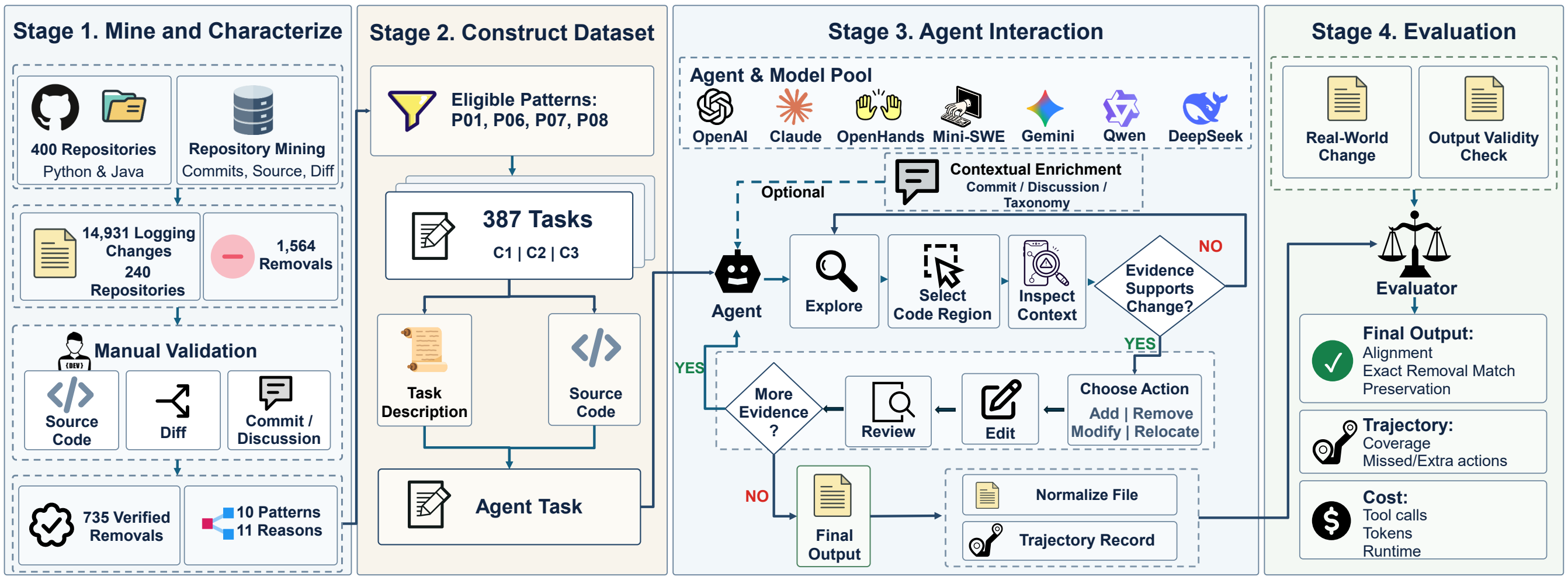

**Figure 2: Overview of the study workflow.**

> Among 14,931 collected logging changes, 1,564 (10.5%) are removals. This shows that removal is a recurring part of logging maintenance and motivates us to examine **how** developers remove logging code and **why** they make these changes.

## 3.2 Validation and Characterization of Logging Removals

Automatically identified removals show where logging code disappears, but they cannot explain how and why it occurs. We therefore manually inspect the 1,564 removal cases using the source code before and after the change, the code diff, commit messages, and pull-request or issue discussions when available. This process results in 735 verified logging removal cases from 164 repositories. We characterize each case in two dimensions: the *removal pattern* describes **how** logging code is removed, while the *removal reason* describes **why** the developer makes the decision. Each case is assigned one primary removal pattern, and one primary removal reason. When the evidence is insufficient to determine developer intent, we assign "Insufficient evidence of intent" rather than inferring a reason. Two authors independently annotate the verified changes. Following prior studies [24, 58, 66], we use Cohen's $\kappa$ [44] to measure inter-annotator agreement. The resulting $\kappa$ values are 0.739 for removal patterns and 0.699 for removal reasons, indicating substantial agreement. The two annotators then discuss disagreements and resolve them whenever possible. If they cannot reach an agreement, a third author is consulted to make the final decision.

Table 1 presents the 10 removal patterns identified from the 735 verified cases. Direct logging statement removal is the most common pattern, accounting for 279 cases (38.0%). However, many removals involve broader program changes, including removal with surrounding functionality, logging infrastructure cleanup, replacement by another mechanism, and relocation. In addition, 149 cases (20.3%) combine multiple removal mechanisms. Table 2 summarizes the 11 removal reasons. Simplification or refactoring is the most common, accounting for 227 cases (30.9%), followed by changed product requirements with 128 cases (17.4%) and redesigned error handling with 90 cases (12.2%). Other reasons include reducing noise or duplication, protecting sensitive information, reducing resource overhead, removing temporary diagnostics, migrating to another output mechanism, and cleaning up logging dependencies or configuration. For 59 cases (8.0%), the available evidence is insufficient to determine developer intent reliably. Overall, these results show that logging removal varies in both how it is performed and why it occurs in real-world changes, ranging from isolated statement removal to changes involving supporting infrastructure or replacement behavior. We use the validated removals to construct our dataset in the next section.

> Direct logging statement removal is the most common pattern, but many logging removals are not limited to deleting individual statements. They may also involve logging infrastructure, replacement with another behavior, or removal together with broader program changes.

# 4 Dataset Construction

In Section 3, we find that logging removal occurs in different scenarios. Building on these findings, we construct our dataset, **LogRem**,

**Table 1: Taxonomy of Logging Removal Patterns**

| Code | Removal Pattern | Definition | Records |
|---|---|---|---|
| **P01** | Direct logging statement removal | Only runtime logging statements are removed while their surrounding executable structures remain. | **279** |
| **P02** | Control-flow block co-deletion | Removed with a complete control-flow block, such as if/else, loop, or switch/case. | **21** |
| **P03** | Exception-handling block co-deletion | Removed together with a complete try/catch, except, finally, or exception-handling branch. | **45** |
| **P04** | Method/class/large implementation co-deletion | Logging disappears as part of the removal of a complete method, class, file, or large implementation block. | **69** |
| **P05** | Feature/parameter/data-path co-deletion | Removed together with a feature, parameter, variable, dependency, execution path, or data path. | **58** |
| **P06** | Logging infrastructure removal | Logger imports, fields, initializers, configurations, appenders, or other logging infrastructure are removed. | **49** |
| **P07** | Replacement by exception/error propagation | Replaced by raise, throw, assert, exception propagation, or an equivalent error contract. | **46** |
| **P08** | Replacement by non-logging output/reporting | Replaced by print, standard output or error, warnings, UI output, callbacks, reports, or other non-logging channels. | **13** |
| **P09** | Logging relocation/centralization | Logging responsibility is moved to another layer, helper, shared logger, or observability component and removed from its original location. | **6** |
| **P10** | Mixed mechanisms | A file-level change contains two or more distinct logging-removal mechanisms that cannot be represented by one primary pattern. | **149** |
| | | **Total** | **735** |

**Table 2: Taxonomy of Logging Removal Reasons**

| Code | Removal Reason | Description | Records |
|---|---|---|---|
| **R01** | Retire obsolete behavior | Associated behavior is obsolete or deprecated. | **33** |
| **R02** | Simplify or refactor implementation | Removed as part of implementation simplification or refactoring. | **227** |
| **R03** | Redesign error handling | The error-handling strategy is redesigned. | **90** |
| **R04** | Reduce noise or duplication | Redundant, excessive, or unnecessary logging output is removed. | **59** |
| **R05** | Protect sensitive information | Logging is removed to prevent exposure of sensitive information. | **17** |
| **R06** | Reduce runtime/resource overhead | Logging is removed to reduce runtime, storage, or other resource overhead. | **29** |
| **R07** | Remove temporary/test diagnostics | Temporary debugging, testing, or diagnostic logging is removed. | **20** |
| **R08** | Migrate observability/output mechanism | Observability or output responsibilities are migrated to another mechanism. | **25** |
| **R09** | Clean up dependency/API plumbing | Removed as part of logger dependency, API, import, configuration, or plumbing cleanup. | **48** |
| **R10** | Implement changed product requirement | Logging is removed because of a feature, requirement, or behavioral change. | **128** |
| **R11** | Insufficient evidence of intent | The evidence is insufficient to infer the developer's removal intent reliably. | **59** |
| | | **Total** | **735** |

from real-world changes in which logging removal can be examined independently from broader program changes. This allows us to study agent behavior without including cases where logging removal is inseparable from larger program changes. The resulting dataset contains 387 Python and Java cases, each with the source code, task description, and real-world change reference.

## 4.1 Selecting Logging Removal Tasks

As shown in Stage 2 of Figure 2, we construct the dataset from the 735 verified logging removal cases characterized in Section 3. We retain cases in which logging removal can be examined independently, including direct removal of logging statements (P01), removal of logging infrastructure such as imports, declarations, initialization, or configuration (P06), and replacement of logging by exception handling or another reporting mechanism (P07, P08). We exclude P02 to P05 because logging is removed together with broader program logic, P09 because logging is relocated rather than removed, and P10 because it combines multiple removal mechanisms. This selection results in 387 cases: 279 direct logging removals (C1), 49 logging infrastructure removals (C2), and 59 logging replacements (C3), as shown in Table 3. These cases form the dataset used in the coding agent study.

## 4.2 Constructing Task Inputs

For each selected case, we manually construct a natural language task description that captures the objective of the original real-world change without revealing the code changes, the logging code removed, or their locations. The description states that the task involves logging removal and that at least one logging-related code should be removed, but does not specify which code to remove, how many removals are needed, or whether infrastructure cleanup or replacement behavior is required. Each task provides this description together with the source file, and the agent must determine the removal scope from the available information. Under the default input setting, only the task description and source code are provided. We also study additional input settings, including Taxonomy guidance, commit messages, and available pull-request or issue discussions. We discuss more details in Section 5.2.

**Table 3: Logging removal task categories derived from the logging removal taxonomy**

| Type | Task Category | Patterns | Change Type | Tasks |
|---|---|---|---|---|
| C1 | Direct logging statement removal | P01 | Remove the intended runtime logging statements while preserving the surrounding program logic. | 279 |
| C2 | Logging infrastructure removal | P06 | Remove logging imports, declarations, initialization, configuration, or other logging infrastructure that is no longer needed. | 49 |
| C3 | Logging replacement | P07, P08 | Remove the logging code and introduce the replacement behavior required by the maintenance change. | 59 |
| | | | **Total** | **387** |

## 4.3 Original Real-World Changes and Output Validity

For each task, we retain the original real-world change as a reference for comparison and apply output validity checks during evaluation. These changes were accepted and merged by developers into the corresponding projects, providing concrete examples of logging removal decisions made in practice. We do not assume that they represent the only valid solution for a task. Using the language specific analysis described in Section 3, we identify the logging code removed in the real-world changes and treat identical logging code at different source locations as separate code. These code form the reference set used in the subsequent comparison with agent generated changes. We also check whether the agent returns a valid final output. An output is considered valid when the required revised source code is present and can be parsed by the corresponding language checker. As shown in Stage 4 of Figure 2, the original real-world change and output validity checks are used to evaluate the agent's output.

# 5 Experiment Setup

As shown in Figure 2, Stage 3 and Stage 4 summarize the agent interaction and evaluation procedure. This section describes the agents and models, the additional task context, the execution settings, and the metrics used to assess final outputs, agent trajectories, and execution cost.

## 5.1 Agents and Models

We evaluate four agents with different model settings. The two commercial agents are OpenAI Codex [47] with *gpt-5.6-luna* [48] and Claude Code [10] with *claude-sonnet-4-6* [2]. We also evaluate two open-source agents: OpenHands [7, 49, 60] and mini-SWE-agent [56], which builds on SWE-agent [64] and is commonly used. For each of these two open source agents, we evaluate them with three backbone LLMs: *DeepSeek-V4-Flash* [13], *Qwen3.8-Flash* [52], and *Gemini 3.7 Flash* [21]. This results in eight experimental settings in total, including two commercial agent and six open-source agent experimental settings.

## 5.2 Task Input Context

The **Base** input context provides only the task description and source code. We examine two additional context sources. **Commit Message and Discussion (CD)** provides task specific historical context from the commit message and available pull-request or issue discussion. **Taxonomy Guidance (T)** provides general guidance derived from the logging removal patterns and reasons with general examples. **Combined Context (CD+T)** provides both CD and T, allowing us to examine if combining the two context sources further improves agent performance. We study four input contexts in total: **Base**, **Base+CD**, **Base+T**, and **Base+CD+T**.

## 5.3 Execution Settings

Each agent and model is evaluated on the same set of dataset cases using the same evaluation procedure and the reasoning effort is set to *medium* for all supported models. The turn limit follows the OpenHands default of 100 iterations [49] and is comparable to recent coding agent evaluations [7]. Each execution ends when the agent returns the final output or reaches the turn limit. We record the final output, agent actions, tool/exploration calls, token usage, and runtime.

## 5.4 Evaluation Metrics

We evaluate both the agent's final output and its trajectory. For the final output, we measure output validity, alignment with real-world changes, and preservation of unrelated code. For the trajectory, we examine removal actions, tool use, token usage, and runtime. The original real-world changes were accepted and merged into their corresponding projects, so we use them as practical references for comparison rather than as the only valid or optimal solutions. The following metrics measure alignment with the real-world changes rather than absolute correctness. For task $i$, let $D_i$ denote the set of logging code removed in the real-world change and $A_i$ those removed by the agent.

**Output Validity and Real-World Change Alignment.** Let $V_i = 1$ when the agent returns a valid final output for task $i$, and $V_i = 0$ otherwise. An output is considered valid when the required revised source code is present and can be processed by the corresponding language checker. A task is aligned with the real-world change when the output is valid, removes exactly the same logging code as in the corresponding real-world change, and leaves unrelated code unchanged. We compute *Output Validity Rate* and *Real-World Change Alignment (RWCA)* Rate as:

$$\text{OutputValidityRate} = \frac{\sum_{i=1}^{N} V_i}{N}, \quad L_i = \mathbb{I}[V_i = 1 \wedge A_i = D_i \wedge U_i = 1] \,. \tag{1}$$

$$\text{RealWorldChangeAlignment(RWCA)} = \frac{\sum_{i=1}^{N} L_i}{N}. \tag{2}$$

**Logging Removal Alignment.** For each task, we compare the logging code removed by the agent with those removed in the real-world change. Let $TP_i$ denote code removed by both, $FN_i$ code removed in real-world change but retained by the agent, and $FP_i$ additional code removed by the agent. We compute Removal Precision as $\frac{TP_i}{TP_i+FP_i}$, Recall as $\frac{TP_i}{TP_i+FN_i}$, and F1 as $\frac{2\,\text{Precision}_i\,\text{Recall}_i}{\text{Precision}_i+\text{Recall}_i}$. We also report *Exact Removal Match*, which measures the percentage of tasks for which the agent removes exactly the same logging code as the corresponding real-world change:

$$\text{ExactRemovalMatch} = \frac{\#\{i \mid A_i = D_i\}}{N}. \quad (3)$$

To characterize differences in removal compared with the real-world change, we report **Under Removal** when logging code removed in the real-world change remains in the agent output, and **Over Removal** when the agent removes logging code retained in the real-world change. For both, we report the number of affected logging code and the percentage of tasks in which they occur.

**Unrelated Code Preservation.** We identify unrelated code changes deterministically from the line-level diff between the source code and the agent output. We exclude logging removals edit marked by the real-world references and edits identified by our rules as logging related. Any remaining non-logging code edit is treated as an unrelated code change. Let $U_i = 1$ if the agent makes no changes to such code in task $i$, and $U_i = 0$ otherwise. The *Unrelated Code Preservation Rate (UCP)* is:

$$\text{UnrelatedCodePreservationRate(UCP)} = \frac{\sum_{i=1}^{N} U_i}{N}. \quad (4)$$

This metric captures changes outside the removal scope that Removal Precision, Recall, and F1 do not reflect. An agent may remove the same logging code as the reference change while also modifying unrelated code. A higher UCP indicates that the agent is less likely to modify unrelated code.

**Agent Trajectory Analysis.** The final code does not reveal how the agent reaches its result, so we analyze the removal actions recorded in the trajectory. For each task, we one-to-one match these actions to the logging code removed in the corresponding real-world change. *Removal Action Coverage* measures the percentage of logging code removed in real-world change that are matched by agent removal actions. *Missed Removal Action Rate* measures the percentage of logging code removed in real-world change with no corresponding agent removal action. *Extra Removal Action Rate* measures the percentage of agent actions that do not match any logging code removed in the corresponding real-world change.

**Tool Use and Execution Cost.** We record tool calls, token usage, and runtime for each task. Exploration calls are separately counted when agents read, search, or inspect source code. Token usage includes input, output, and total tokens, while runtime is measured from the start of the task until the agent returns the revised source code or reaches the turn limit. We report average cost per task and two alignment normalized measures. *Tokens per Aligned Task* (Tok./Align.) divides total token usage by the number of tasks satisfying the Real-World Change Alignment criterion, while *Runtime per Aligned Task* (Time/Align.) is computed analogously using total runtime.

## 6 Results

We evaluate coding agents on LogRem through the following research questions:

- **RQ1**: How do coding agent outputs compare with real-world logging removal changes? (Final Output Alignment)
- **RQ2**: How do coding agents behave during logging removal tasks? (Agent Trajectory Analysis)
- **RQ3**: How do execution costs differ across coding agent settings? (Execution Cost)
- **RQ4**: How do different forms of context affect results? (Context Effects)

### 6.1 RQ1: How do coding agent outputs compare with real-world logging removal changes?

***Motivation.*** Removing logging code often requires more than deleting individual logging statements. Developers may need to inspect the surrounding implementation, dependencies, and maintenance context to decide which logging code should be removed and whether infrastructure cleanup or replacement is needed. RQ1 examines how closely current coding agents can reproduce such changes made in real-world projects.

***Approach.*** We evaluate all model settings on the 387 logging removal tasks under the `Base` input context. For each task, we compare the agent output with the corresponding real-world change. We first examine overall alignment across the eight settings, then analyze differences in removal scope through under and over removals. We further compare results across C1, C2, and C3 to determine whether different logging removal categories exhibit distinct output characteristics.

***Results.*** **Coding agents exhibit limited alignment with real-world logging removal decisions under the `Base` input context.** As shown in Table 4, RWCA rate ranges from 11.1% to 19.6% across the eight settings. OpenHands with Gemini 3.7 Flash reaches 19.6%, while Claude Code with Sonnet 4.6 reaches 11.1%. This indicates a substantial gap between removals generated by coding agents and the changes made in real-world commits.

**Agents usually produce valid code, but often make logging changes that differ from the corresponding real-world changes.** Output Validity Rates range from 95.6% to 100.0%, showing that most agents return source code that passes the basic language checks. However, Exact Removal Match reaches at most 35.9%, and UCP ranges from 42.1% to 59.9%. Thus, producing valid code does not mean that the agent removes the intended logging code or preserves the surrounding code. Agents often remove different logging code from the real-world changes or make extra changes to unrelated code.

**Performance varies across agent and model settings, and no setting performs consistently well across all aspects of logging removal.** Codex with GPT-5.6 Luna achieves the highest Unrelated Code Preservation Rate at 59.9% and the highest Removal Precision at 56.9%, but misses at least one removal from real-world changes in 54.0% of its tasks. OpenHands with DeepSeek V4 Flash reaches the highest Removal Recall at 67.4%, but makes additional removals in 61.5% of its tasks. Claude Code with Sonnet 4.6 shows the highest rate of additional removals at 64.9%. These differences show that current agent and model combinations handle logging removal unevenly. A setting that performs better at identifying intended removals or preserving unrelated code may still miss other removals or remove additional logging code.

**Table 4: Final output metrics across agent and model settings.**

| Setting | RWCA (%) | Exact Removal Match (%) | UCP (%) | Output Validity (%) | F1 (%) | Precision (%) | Recall (%) |
|---|---|---|---|---|---|---|---|
| Codex + GPT-5.6 Luna | 18.3 | 30.7 | **59.9** | 99.7 | **49.5** | **56.9** | **53.1** |
| Claude Code + Sonnet 4.6 | **11.1** | **26.1** | **42.1** | 100.0 | 50.6 | **49.2** | 65.0 |
| Mini-SWE + DeepSeek V4 Flash | 18.1 | 31.5 | 45.5 | 99.7 | 53.5 | 54.5 | 64.9 |
| Mini-SWE + Gemini 3.7 Flash | 18.3 | 33.3 | 45.7 | 95.6 | 54.2 | 55.0 | 65.6 |
| Mini-SWE + Qwen 3.8 Flash | 16.8 | 31.5 | 48.6 | 100.0 | 53.6 | 54.5 | 65.0 |
| OpenHands + DeepSeek V4 Flash | 13.4 | 27.4 | 43.4 | 100.0 | 52.2 | 51.2 | **67.4** |
| OpenHands + Gemini 3.7 Flash | **19.6** | **35.9** | 48.3 | 100.0 | **55.5** | 56.7 | 65.6 |
| OpenHands + Qwen 3.8 Flash | 15.2 | 30.7 | 45.2 | 100.0 | 51.7 | 52.7 | 61.4 |

**Table 5: Over removal and under removal across agent and model settings.**

| Setting | Over Removal (code) | Tasks with Over Removal (%) | Under Removal (code) | Tasks with Under Removal (%) |
|---|---|---|---|---|
| Codex + GPT-5.6 Luna | **291** | **49.9** | **514** | **54.0** |
| Claude Code + Sonnet 4.6 | 726 | **64.9** | 357 | 41.1 |
| Mini-SWE + DeepSeek V4 Flash | 713 | 53.7 | 339 | 42.6 |
| Mini-SWE + Gemini 3.7 Flash | 593 | 50.4 | 356 | **39.3** |
| Mini-SWE + Qwen 3.8 Flash | 750 | 56.6 | 370 | 41.9 |
| OpenHands + DeepSeek V4 Flash | **813** | 61.5 | **338** | **39.3** |
| OpenHands + Gemini 3.7 Flash | 490 | 50.6 | 359 | 40.1 |
| OpenHands + Qwen 3.8 Flash | 685 | 57.6 | 394 | 45.0 |

**Table 6: Final output metrics by logging removal task category, averaged across agent and model settings.**

| Task Category | RWCA (%) | Exact RM Match (%) | UCP (%) | Precision (%) | Recall (%) | Over RM (Code/task) | Under RM (Code/task) |
|---|---|---|---|---|---|---|---|
| C1 | **19.4** | 28.7 | **52.0** | 51.2 | 65.8 | 1.82 | **0.69** |
| C2 | 11.7 | **62.8** | **25.5** | **92.1** | **79.9** | **0.27** | 1.22 |
| C3 | **5.9** | **14.8** | 43.6 | **34.9** | **39.1** | **1.89** | **2.17** |

**Alignment with real-world changes varies substantially across different logging removal categories.** As shown in Table 6, C1 direct logging statement removal reaches the highest overall alignment, with a RWCA Rate of 19.4% and an Unrelated Code Preservation Rate of 52.0%. C2 logging infrastructure removal shows much stronger alignment on the logging code removed in real-world changes, reaching 62.8% Exact Removal Match, 79.9% Removal Recall, and 92.1% Removal Precision. However, its RWCA Rate is only 11.7%, mainly because agents often modify code beyond the logging code removed in real-world changes, as reflected by its 25.5% Unrelated Code Preservation Rate. C3 logging replacement shows the lowest alignment across most measures, with 5.9% RWCA, 14.8% Exact Removal Match, 39.1% Removal Recall, and 34.9% Removal Precision. It also has the largest numbers of missed and additional removals, averaging 2.17 and 1.89 logging code per task. These results show that alignment with real-world changes depends not only on removing the same logging code, but also on preserving code that remains untouched in the corresponding real-world changes. The source of disagreement also differs across categories, with C2 showing relatively strong agreement on removals but weak overall alignment, while C3 differs more broadly from the real-world changes.

> **RQ1 Summary:** Coding agents usually return valid code but show limited alignment with real-world logging removals. Under `Base` input setting, RWCA ranges from 11.1% to 19.6% despite Output Validity Rates of 95.6% to 100.0%, and no setting performs consistently well across all measures. Alignment also varies by task type, with C1 reaching 19.4% RWCA, C2 often matching the removed logging code but changing other code, and C3 reaching only 5.9%. Overall, the gap between validity and alignment points to two recurring challenges: identifying which logging code to remove and limiting changes to the intended scope.

## 6.2 RQ2: How do coding agents behave during logging removal tasks?

***Motivation.*** RQ1 compares the final agent output with the real-world change, but the final code does not show how the agent arrived at that result. An agent may miss logging code removed in the corresponding real-world change or remove additional code during execution, even if some of these actions are later revised. RQ2 therefore examines the removal actions recorded in agent trajectories to understand how agents select what to remove and how this behavior differs across agent settings and task categories.
***Approach.*** Building on the experimental setup of RQ1, we further analyze the removal actions recorded in agent trajectories. We extract logging removal actions from the recorded code edits and match them to the logging code removed in the corresponding real-world change. The matching procedure is deterministic and does not rely on an LLM judge. Each agent removal action is matched one-to-one with a real-world removal based on the affected logging code and its source location, so the same real-world removal cannot be counted multiple times. Unmatched real-world removals are treated as missed actions, while unmatched agent removals are treated as extra actions. We first compare removal action behavior across agent and model settings, and then aggregate the results across C1, C2, and C3 to examine differences among logging removal categories.

Table 7: Trajectory Level Removal Action Behavior of Different Agents and Models

| Agent + Model | Removal Action Coverage (%) | Missed Removal Action (%) | Extra Removal Action (%) |
|---|---|---|---|
| Codex + GPT-5.6 Luna | **51.8** | **48.2** | 37.3 |
| Claude Code + Sonnet 4.6 | 62.7 | 37.3 | **43.6** |
| Mini-SWE + DeepSeek V4 Flash | 63.6 | 36.4 | 35.7 |
| Mini-SWE + Gemini 3.7 Flash | 63.0 | 37.0 | **34.6** |
| Mini-SWE + Qwen 3.8 Flash | 63.9 | 36.1 | 39.0 |
| OpenHands + DeepSeek V4 Flash | **65.8** | **34.2** | 40.3 |
| OpenHands + Gemini 3.7 Flash | 64.2 | 35.8 | 35.8 |
| OpenHands + Qwen 3.8 Flash | 58.8 | 41.2 | 40.4 |

Table 8: Trajectory Level Removal Action Behavior by Logging Removal Category

| Task Category | Removal Action Coverage (%) | Missed Removal Action (%) | Extra Removal Action (%) |
|---|---|---|---|
| C1: Direct logging statement removal | 65.6 | 34.4 | 40.6 |
| C2: Logging-infrastructure cleanup | **76.6** | **23.4** | **5.3** |
| C3: Replacement-based removal | **30.9** | **69.1** | **55.0** |

***Results.*** **Agent trajectories show that incomplete and additional removal decisions occur throughout the process of completing the task.** As shown in Table 7, Removal Action Coverage ranges from 51.8% to 65.8% across the eight settings, meaning that a substantial portion of the logging code removed in real-world changes are never targeted by the agents. At the same time, Extra Removal Action Rates range from 34.6% to 43.6%, showing that additional removal actions are common across all settings. OpenHands with DeepSeek V4 Flash reaches the highest coverage at 65.8%, but 40.3% of its removal actions are additional. Mini-SWE with Gemini 3.7 Flash has the lowest Extra Removal Action Rate at 34.6%, while covering 63.0% of real-world removals. These results show that no setting consistently identifies real-world removal targets while avoiding additional removal actions.

**Agent removal behavior varies substantially across logging removal categories.** As shown in Table 8, C2 logging infrastructure removal has the closest action-level correspondence with real-world changes, reaching 76.6% Removal Action Coverage with only 5.3% Extra Removal Actions. This helps explain the RQ1 results more precisely. Agents generally identify the logging code to remove in C2, while much of the remaining disagreement in the final output comes from changes elsewhere in the code. In contrast, C3 logging replacement shows divergence during the removal process itself, with only 30.9% coverage, 69.1% missed actions, and 55.0% extra actions. C1 lies between these two cases, with 65.6% coverage and 40.6% extra actions. These results show that trajectory analysis can distinguish whether disagreement arises while choosing logging code to remove or from other edits made around those removals.

> **RQ2 Summary:** Agent trajectories show that divergence from real-world changes begins during removal selection. Across settings, agents target only 51.8% to 65.8% of real-world removals and frequently perform additional actions. This behavior also varies by task category, with C2 showing close agreement on removal targets and C3 showing substantial divergence during selection.

## 6.3 RQ3: How do execution costs differ across coding agent settings?

***Motivation.*** Agent settings differ in the changes they produce, as well as in the computational effort required to produce them. Evaluating alignment alone therefore does not show how efficiently a model setting reaches its result. RQ3 examines how token usage, runtime, and tool interaction vary across model settings and whether more resource usage is associated with closer alignment with real-world changes.

***Approach.*** We examine execution costs, including token usage, runtime, exploration calls, and tool calls, based on the agent runs evaluated in RQ1. We also report Tokens per Aligned Task and Runtime per Aligned Task, which divide the total resource consumption across all tasks by the number of tasks that satisfy the Real-World Change Alignment (RWCA) criterion.

***Results.*** **Execution cost varies substantially across model settings.** As shown in Table 9, Claude Code with Sonnet 4.6 uses the fewest total tokens at 57.2K per task and has the lowest runtime at 48.2 seconds per task. In contrast, Mini-SWE with DeepSeek V4 Flash consumes 1.10M tokens per task and performs 36.2 tool calls on average, while OpenHands with Gemini 3.7 Flash has the longest runtime at 357.4 seconds per task. The cost per aligned task also varies substantially, from 515.1K to 6.07M tokens and from 273.8 to 1,819.7 seconds.

**Table 9: Execution Cost across Agent and Model Settings**

| Agent + Model | Input Tok./Task (K) | Output Tok./Task (K) | Total Tok./Task (K) | Runtime/Task (s) | Exploration Calls | Tool Calls | Tok./Align(K) | Time/Align(s) |
|---|---|---|---|---|---|---|---|---|
| Codex + GPT 5.6 Luna | 129.7 | **1.6** | 131.3 | 50.2 | 3.6 | **5.5** | 715.7 | **273.8** |
| Claude Code + Sonnet 4.6 | **54.8** | 2.4 | **57.2** | **48.2** | 3.2 | 5.8 | **515.1** | 434.2 |
| Mini-SWE + DeepSeek V4 Flash | **1066.9** | **30.9** | **1097.8** | 256.9 | 8.7 | **36.2** | **6069.3** | 1420.1 |
| Mini-SWE + Gemini 3.7 Flash | 422.5 | 11.9 | 434.3 | 64.6 | **1.4** | 25.9 | 2367.5 | 352.1 |
| Mini-SWE + Qwen 3.8 Flash | 442.3 | 16.2 | 458.4 | 222.2 | 8.1 | 32.7 | 2729.4 | 1322.7 |
| OpenHands + DeepSeek V4 Flash | 280.4 | 14.5 | 294.9 | 140.8 | **9.1** | 12.3 | 2194.7 | 1047.7 |
| OpenHands + Gemini 3.7 Flash | 399.8 | 13.8 | 413.6 | **357.4** | 9.0 | 16.8 | 2106.1 | **1819.7** |
| OpenHands + Qwen 3.8 Flash | 157.0 | 8.1 | 165.0 | 129.4 | 8.5 | 11.7 | 1082.5 | 848.6 |

**Table 10: Effect of optional task context on Claude Code with Sonnet 4.6.**

| Setting | RWCA (%) | Exact Removal Match (%) | UCP (%) | Tokens/Task (K) | Runtime/Task (s) | Tok./Align (K) | Time/Align (s) |
|---|---|---|---|---|---|---|---|
| Base | 11.1 | 26.1 | 42.1 | 57.2 | 48.2 | 515.1 | 434.2 |
| Base+T | 12.7 (↑1.6) | 27.1 (↑1.0) | 44.7 (↑2.6) | 53.2 (↓4.1) | 42.6 (↓5.7) | 420.0 (↓95.1) | 336.1 (↓98.1) |
| Base+CD | **22.2** (↑11.1) | **42.4** (↑16.3) | 46.3 (↑4.1) | 54.8(↓2.4) | 38.2 (↓10.0) | **246.6** (↓268.5) | **172.0** (↓262.2) |
| Base+CD+T | 21.4 (↑10.3) | 41.9 (↑15.8) | **47.3** (↑5.2) | 58.1 (↑0.8) | **37.9** (↓10.3) | 270.8 (↓244.3) | 176.8 (↓257.4) |

**Model settings with comparable alignment to real-world changes can require substantially different execution costs.** Codex with GPT 5.6 Luna reaches an 18.3% RWCA Rate using 131.3K tokens and 50.2 seconds per task. Mini-SWE with Gemini 3.7 Flash reaches the same 18.3% alignment rate but consumes 434.3K tokens and 64.6 seconds per task. Mini-SWE with DeepSeek V4 Flash reaches a similar alignment rate of 18.1% while consuming 1.10M tokens and 256.9 seconds per task. Thus, settings with nearly identical alignment rates differ by more than eight times in token usage and more than five times in runtime.

**Greater execution cost does not necessarily lead to closer alignment with real-world changes.** OpenHands with Gemini 3.7 Flash reaches the highest RWCA Rate at 19.6%, but requires 413.6K tokens and 357.4 seconds per task. Its cost per aligned task reaches 2.11M tokens and 1,819.7 seconds. In comparison, Codex with GPT 5.6 Luna reaches an 18.3% alignment rate with 715.7K tokens and 273.8 seconds per aligned task. These results show that greater resource consumption does not consistently correspond to higher alignment with real-world changes.

> **RQ3 Summary:** Execution cost varies substantially across model settings and does not consistently correspond to higher alignment. Model settings with similar RWCA can differ by several times in resource use; for example, two settings with 18.3% RWCA consume 131.3K and 434.3K tokens per task, respectively. Both resource consumption and output alignment should be considered when comparing coding agent settings.

## 6.4 RQ4: How do different forms of context affect results?

***Motivation.*** In the previous RQs, the `Base` input setting provides only the task description and source code. In practice, additional context may be available to provide different forms of information about the task. Commit messages and developer discussions provide task specific historical information, whereas the taxonomy may provide general guidance derived from recurring logging removal patterns and reasons. RQ4 examines how these context sources affect alignment with real-world changes and execution cost.

***Approach.*** We conduct a focused context comparison using two agent and model settings selected from the results in RQ1 with the lowest and highest RWCA: Claude Code with Sonnet 4.6 (11.1% RWCA) and OpenHands with Gemini 3.7 Flash (19.6% RWCA). For each setting, we evaluate the four input settings: `Base`, `Base+T`, `Base+CD`, and `Base+CD+T`. All other task and execution settings are held constant. We compare differences in output alignment and execution cost across these context settings to examine the effects of historical context, taxonomy guidance, and their combination.

***Results.*** **Task specific historical context produces the largest improvement in alignment with real-world changes.** As shown in Tables 10 and 11, `Base+CD` increases RWCA for Claude Code with Sonnet 4.6 from 11.1% to 22.2% and increases Exact Removal Match from 26.1% to 42.4%. For OpenHands with Gemini 3.7 Flash, RWCA increases from 19.6% to 28.2%, while Exact Removal Match increases from 35.9% to 49.9%. Among the four input settings, `Base+CD` reaches the highest RWCA and Exact Removal Match for both agent and model settings. It also lowers the cost per aligned task. Tokens per Aligned Task decrease from 515.1K to 246.6K for Claude Code and from 2.11M to 1.47M for OpenHands, while Runtime per Aligned Task decreases from 434.2 to 172.0 seconds and from 1819.7 to 1125.3 seconds, respectively.

**Taxonomy guidance helps improve alignment and consistently reduces runtime.** For Claude Code, `Base+T` increases RWCA from 11.1% to 12.7%, Exact Removal Match from 26.1% to 27.1%, and Unrelated Code Preservation from 42.1% to 44.7%. It also reduces token usage and runtime per task. For OpenHands, RWCA increases from 19.6% to 20.9%, Exact Removal Match from 35.9% to 36.2%, and Unrelated Code Preservation from 48.3% to 50.9%. Runtime decreases from 48.2 to 42.6 seconds for Claude Code and from 357.4 to 230.9 seconds for OpenHands. These results show that taxonomy guidance increased alignment and reduced mean runtime in both evaluated settings. One possible explanation is that the taxonomy provides reusable guidance about common removal patterns and reasons, which may help agents narrow the plausible removal scope and reach decisions more quickly.

**Table 11: Effect of Optional Task Context on OpenHands with Gemini 3.7 Flash.**

| Setting | RWCA(%) | Exact Removal Match(%) | UCP(%) | Tokens/Task(K) | Runtime/Task(s) | Tok./Align(K) | Time/Align(s) |
|---|---|---|---|---|---|---|---|
| Base | 19.6 | 35.9 | 48.3 | **413.6** | 357.4 | 2106.1 | 1819.7 |
| Base+T | 20.9 (↑1.3) | 36.2 (↑0.3) | 50.9 (↑2.6) | 468.5 (↑54.9) | 230.9 (↓126.5) | 2238.3 (↑132.2) | 1103.1 (↓716.6) |
| Base+CD | **28.2** (↑8.5) | **49.9** (↑14.0) | 48.3 | 412.7 (↓0.9) | 316.9 (↓40.4) | **1465.3** (↓640.8) | 1125.3 (↓694.4) |
| Base+CD+T | 26.4 (↑6.8) | 48.3 (↑12.4) | 49.4 (↑1.0) | 455.2 (↑41.6) | **202.0** (↓155.4) | 1727.2 (↓378.9) | **766.3** (↓1053.5) |

**When historical context is already available, taxonomy guidance mainly provides additional runtime improvement.** Adding taxonomy guidance to `Base+CD` does not further improve alignment. For Claude Code, RWCA changes from 22.2% to 21.4%, while Exact Removal Match changes from 42.4% to 41.9%. For OpenHands, the corresponding values change from 28.2% to 26.4% and from 49.9% to 48.3%. However, `Base+CD+T` achieves the lowest runtime among all four input settings for both agents, reaching 37.9 seconds for Claude Code and 202.0 seconds for OpenHands. These results indicate that the two context sources provide different benefits. Historical context contributes more to alignment with real-world changes, while taxonomy guidance provides a more consistent reduction in runtime.

> **RQ4 Summary:** Task specific historical context provides the largest alignment improvements, increasing RWCA from 11.1% to 22.2% for Claude Code and from 19.6% to 28.2% for OpenHands. Taxonomy guidance consistently improves efficiency by reducing runtime. When both are available, historical context remains more effective for alignment, while taxonomy guidance provides additional runtime benefits.

# 7 Discussion

## 7.1 Case Study

Figure 3 compares the results of two agents on the same task under `Base` input setting. For readability, the figure shows a shortened task description and code snippet. The developer discussion is included only to show the considerations raised during code review.

The developer removes the warning emitted when no recognized Conscrypt provider is found while preserving the subsequent `return null`. The associated developer discussion notes that the absence of Conscrypt is normal on the JDK, where another ALPN implementation may be used, whereas removing the default Android security providers is considered extremely rare. Claude Code with Sonnet 4.6 leaves this warning unchanged and instead removes logging from the Android version checks, introducing extra logging changes beyond the real-world removal. OpenHands with Gemini 3.7 Flash removes the warning while preserving the surrounding provider selection logic and the `Level.FINE` logs shown in the figure. Thus, OpenHands matches the real-world removal in this case, whereas Claude Code adopts a different removal scope. This case illustrates that agents can produce valid outputs while making substantially different logging removal decisions. Comparing their outputs with the observed real-world change therefore helps characterize differences in removal scope without treating the real-world change as the only valid solution.

## 7.2 Implications

We discuss the implications of our findings for practitioners and researchers. `IMP-P#` denotes implications for practitioners, and `IMP-R#` denotes implications for researchers.

**IMP-P1. Review logging removal as both a deletion and a preservation task.** Our results show that an agent can identify relevant logging code while still changing code that left untouched in the real-world change. Logging replacement poses a different challenge because the agent must determine both what to remove and what behavior should replace it. In practice, reviewing an agent generated removal should therefore check not only whether the intended logging disappears, but also whether surrounding behavior and remaining logging are preserved and whether required cleanup or replacement is handled correctly. The removal patterns provide a useful way to organize these checks. Direct removal calls for verifying the target statements and surrounding logic, infrastructure removal requires checking imports, declarations, and configuration, and replacement requires validating the new behavior.

**IMP-P2. Use project history when available, and use taxonomy guidance when task-specific evidence is limited.** RQ4 shows that commit messages and developer discussions provide the largest improvements in alignment for both evaluated settings. RWCA rate increases from 11.1% to 22.2% for Claude Code and from 19.6% to 28.2% for OpenHands. Taxonomy guidance produces smaller alignment improvements but consistently reduces runtime. These results suggest a practical ordering for agent context. When available, issue discussions, pull-requests, and commit messages should be used to explain why a logging change is needed. When such evidence is missing or incomplete, the taxonomy can provide general guidance about common removal patterns and reasons. This allows the agent to use project specific evidence when possible while still benefiting from reusable knowledge when that evidence is unavailable.

**IMP-P3. Select agent settings based on both alignment and execution cost.** RQ3 shows that settings with similar alignment can require substantially different computational resources. For example, Codex with GPT-5.6 Luna and Mini-SWE with Gemini 3.7 Flash both reach an 18.3% RWCA Rate, while token usage ranges from 131.3K to 434.3K tokens per task. Mini-SWE with DeepSeek V4 Flash consumes more than one million tokens per task without achieving higher alignment. Practical systems should therefore consider both alignment and cost when selecting agent and model settings rather than assuming that greater resource consumption will improve alignment.

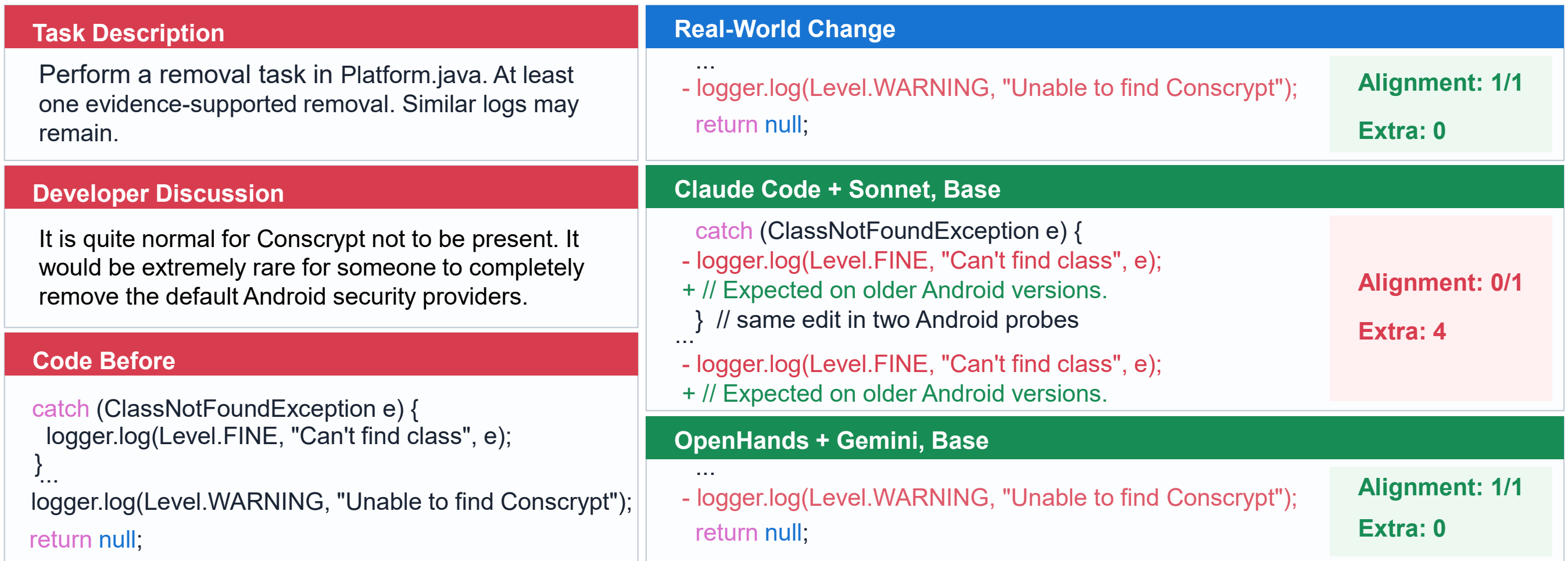

**Figure 3: Different logging removal decisions made by two agents on the same task under Base input setting.**

**IMP-R1. Use removal taxonomies and task categories to structure future logging removal research.** Our study shows that logging removal varies in both how logging code changes and why those changes occur in real-world projects. The pattern and reason taxonomies characterize this broader variation, while C1, C2, and C3 define categories of logging removal that can be examined independently in agent-based logging removal tasks. Our results further show substantial differences across these categories in both final outputs and removal actions. Treating logging removal as one broad category may therefore hide important differences in the required changes and agent behavior. Future research can use these taxonomies to structure dataset construction, evaluation design, and automated techniques for different logging removal categories.

**IMP-R2. Apart from final output, agent evaluation should also study how maintenance decisions are made.** Our trajectory analysis shows that agents can omit intended removals and perform extra removal actions before producing a final patch, and some intermediate actions are later revised. These behaviors cannot be recovered from the final output alone. Future research can use agent trajectories to examine where scope differences emerge, including repository exploration, evidence collection, action selection, edit revision, and stopping decisions. Such analysis may help identify when and why unnecessary scope expansion occurs during the agent's problem-solving process.

## 8 Threats to Validity

**Construct validity.** Real-world changes provide concrete examples of logging removal decisions that were accepted and merged into their projects, but they do not necessarily represent the only valid solution. An agent may produce a different removal scope that is still reasonable under another observability or maintenance objective. We therefore interpret Real-World Change Alignment and related measures as alignment with the real-world change rather than absolute correctness.

**Internal validity.** Dataset construction combines automated logging change identification with manual validation, and annotation, which may be affected by identification errors, ambiguous repository evidence, or subjective judgments. To reduce annotation inconsistency, two authors independently annotate the verified cases and discuss disagreements to reach consensus. When an agreement cannot be reached, a third author is engaged to make the final decision. Cohen's $\kappa$ is 0.739 for removal patterns and 0.699 for removal reasons. We inspect source code, diffs, commit messages, and pull-request or issue discussions when available, and assign an insufficient-evidence category when removal intent cannot be inferred reliably. Task descriptions are constructed without revealing the real-world change or removal locations.

**External validity.** We conduct the study on Python and Java open-source projects. The findings may not directly generalize to other programming languages, logging ecosystems, project domains, or more tightly coupled maintenance changes. Our evaluation covers multiple agent frameworks, models, and removal categories, but future systems and settings may perform differently. As coding agent systems continue to evolve, replication on additional models, repositories, and maintenance settings would further assess the generality of the findings.

## 9 Conclusion

In this paper, we study logging removal as a distinct software maintenance activity. We manually validate real logging removal cases, derive 10 removal patterns and 11 removal reasons, and construct **LogRem** with 387 cases covering three forms of logging removal. Coding agents usually produce valid outputs, but their results show limited alignment with observed real-world changes, and both final outputs and removal trajectories differ substantially across removal categories. Execution cost also varies widely without consistently corresponding to higher alignment. Task specific historical context provides the largest alignment gains, while taxonomy guidance provides reusable context and reduces runtime in the evaluated settings. These findings show that output validity alone does not capture whether an agent makes logging removal decisions consistent with observed real-world changes, and that different removal categories expose distinct agent challenges. Our taxonomies and **LogRem** provide a basis for future logging removal datasets, evaluations, and automated techniques.

## References

[1] Anunay Amar and Peter C. Rigby. 2019. Mining historical test logs to predict bugs and localize faults in the test logs. In *Proceedings of the 41st International Conference on Software Engineering, ICSE 2019, Montreal, QC, Canada, May 25-31, 2019*, Joanne M. Atlee, Tevfik Bultan, and Jon Whittle (Eds.). IEEE / ACM, 140–151. doi:10.1109/ICSE.2019.00031

[2] Anthropic. 2026. Introducing Claude Sonnet 4.6. https://www.anthropic.com/news/claude-sonnet-4-6.

[3] Islem Bouzenia and Michael Pradel. 2025. Understanding software engineering agents: A study of thought-action-result trajectories. In *2025 40th IEEE/ACM International Conference on Automated Software Engineering (ASE)*. IEEE, 2846–2857.

[4] Boyuan Chen and Zhen Ming (Jack) Jiang. 2017. Characterizing logging practices in Java-based open source software projects - a replication study in Apache Software Foundation. *Empir. Softw. Eng.* 22, 1 (2017), 330–374. doi:10.1007/S10664-016-9429-5

[5] Boyuan Chen and Zhen Ming (Jack) Jiang. 2019. Extracting and studying the Logging-Code-Issue- Introducing changes in Java-based large-scale open source software systems. *Empir. Softw. Eng.* 24, 4 (2019), 2285–2322. doi:10.1007/S10664-019-09690-0

[6] Boyuan Chen and Zhen Ming (Jack) Jiang. 2020. Studying the use of Java logging utilities in the wild. In *ICSE '20: 42nd International Conference on Software Engineering, Seoul, South Korea, 27 June - 19 July, 2020*, Gregg Rothermel and Doo-Hwan Bae (Eds.). ACM, 397–408. doi:10.1145/3377811.3380408

[7] Junkai Chen, Huihui Huang, Yunbo Lyu, Junwen An, Jieke Shi, Chengran Yang, Ting Zhang, Haoye Tian, Yikun Li, Zhenhao Li, Xin Zhou, Xing Hu, and David Lo. 2026. Securevibebench: Benchmarking secure vibe coding of ai agents via reconstructing vulnerability-introducing scenarios. In *Proceedings of the 64th Annual Meeting of the Association for Computational Linguistics (Volume 1: Long Papers)*. 24144–24168.

[8] Junkai Chen, Chengran Yang, Xing Hu, Zhenhao Li, Xin Xia, and David Lo. 2026. AgentExecutor: Partial Code Execution via Agentic Context Generation. *arXiv preprint arXiv:2608.05959* (2026).

[9] Zhi Chen and Lingxiao Jiang. 2025. Evaluating software development agents: Patch patterns, code quality, and issue complexity in real-world github scenarios. In *2025 IEEE international conference on software analysis, evolution and reengineering (SANER)*. IEEE, 657–668.

[10] Claude Code. 2026. Claude Code Docs. https://docs.anthropic.com/en/docs/claude-code. Last Access: 2026.

[11] Leuson Da Silva, Altaf Allah Abbassi, Imen Trabelsi, Paulo Borba, and Foutse Khomh. 2026. GitHub Template Repositories: Served Domains, Maintenance, and Practitioner Guidelines. *arXiv preprint arXiv:2606.14616* (2026).

[12] Ozren Dabic, Emad Aghajani, and Gabriele Bavota. 2021. Sampling projects in github for MSR studies. In *2021 IEEE/ACM 18th International Conference on Mining Software Repositories (MSR)*. IEEE, 560–564.

[13] Deepseek. 2026. DeepSeek. https://api-docs.deepseek.com/quick_start/pricing/. Last Access: 2026.

[14] Rui Ding, Hucheng Zhou, Jian-Guang Lou, Hongyu Zhang, Qingwei Lin, Qiang Fu, Dongmei Zhang, and Tao Xie. 2015. Log2: A {Cost-Aware} logging mechanism for performance diagnosis. In *2015 USENIX annual technical conference (USENIX ATC 15)*. 139–150.

[15] Zishuo Ding, Heng Li, and Weiyi Shang. 2022. LoGenText: Automatically Generating Logging Texts Using Neural Machine Translation. In *IEEE International Conference on Software Analysis, Evolution and Reengineering, SANER 2022, Honolulu, HI, USA, March 15-18, 2022*. IEEE, 349–360. doi:10.1109/SANER53432.2022.00051

[16] Zishuo Ding, Yiming Tang, Xiaoyu Cheng, Heng Li, and Weiyi Shang. 2024. *LoGenText-Plus*: Improving Neural Machine Translation Based Logging Texts Generation with Syntactic Templates. *ACM Trans. Softw. Eng. Methodol.* 33, 2 (2024), 38:1–38:45. doi:10.1145/3624740

[17] Zishuo Ding, Yiming Tang, Yang Li, Heng Li, and Weiyi Shang. 2023. On the temporal relations between logging and code. In *2023 IEEE/ACM 45th International Conference on Software Engineering (ICSE)*. IEEE, 843–854.

[18] Patrick Loic Foalem, Foutse Khomh, and Heng Li. 2024. Studying logging practice in machine learning-based applications. *Information and Software Technology* 170 (2024), 107450.

[19] Qiang Fu, Jian-Guang Lou, Qingwei Lin, Rui Ding, Dongmei Zhang, and Tao Xie. 2013. Contextual analysis of program logs for understanding system behaviors. In *Proceedings of the 10th Working Conference on Mining Software Repositories, MSR '13, San Francisco, CA, USA, May 18-19, 2013*, Thomas Zimmermann, Massimiliano Di Penta, and Sunghun Kim (Eds.). IEEE Computer Society, 397–400. doi:10.1109/MSR.2013.6624054

[20] Google. 2023. Four steps to managing your Cloud Logging costs on a budget. https://cloud.google.com/blog/products/devops-sre/cloud-logging-cost-management-best-practices. Accessed: 2026-09-28.

[21] Google. 2026. Introducing Gemini 3.7 Flash. https://blog.google/innovation-and-ai/models-and-research/gemini-models/introducing-gemini-3-7-flash/. Last Access: 2026.

[22] Shenghui Gu, Guoping Rong, He Zhang, and Haifeng Shen. 2022. Logging practices in software engineering: A systematic mapping study. *IEEE transactions on software engineering* 49, 2 (2022), 902–923.

[23] Pengfei He, Zhenwei Dai, Bing He, Hui Liu, Xianfeng Tang, Hanqing Lu, Juanhui Li, Jiayuan Ding, Subhabrata Mukherjee, Suhang Wang, et al. 2026. Traject-bench: a trajectory-aware benchmark for evaluating agentic tool use. In *International Conference on Learning Representations*, Vol. 2026. 61766–61801.

[24] Xing Hu, Xin Xia, David Lo, Zhiyuan Wan, Qiuyuan Chen, and Thomas Zimmermann. 2022. Practitioners' Expectations on Automated Code Comment Generation. In *Proceedings of the 44th international conference on software engineering*. 1693–1705. doi:10.1145/3510003.3510152

[25] Carlos E Jimenez, John Yang, Alexander Wettig, Shunyu Yao, Kexin Pei, Ofir Press, and Karthik Narasimhan. 2023. Swe-bench: Can language models resolve real-world github issues?, 2024. *URL https://arxiv. org/abs/2310.06770* 7 (2023).

[26] Suhas Kabinna, Cor-Paul Bezemer, Weiyi Shang, Mark D. Syer, and Ahmed E. Hassan. 2018. Examining the stability of logging statements. *Empir. Softw. Eng.* 23, 1 (2018), 290–333. doi:10.1007/S10664-017-9518-0

[27] Jinhan Kim, Valeriy V. Savchenko, Kihyuck Shin, Konstantin S. Sorokin, Hyunseok Jeon, Georgiy A. Pankratenko, Sergey Markov, and Chul-Joo Kim. 2020. Automatic abnormal log detection by analyzing log history for providing debugging insight. In *ICSE-SEIP 2020: 42nd International Conference on Software Engineering, Software Engineering in Practice, Seoul, South Korea, 27 June - 19 July, 2020*, Gregg Rothermel and Doo-Hwan Bae (Eds.). ACM, 71–80. doi:10.1145/3377813.3381371

[28] Heng Li, Tse-Hsun Chen, Weiyi Shang, and Ahmed E Hassan. 2018. Studying software logging using topic models. *Empirical Software Engineering* 23, 5 (2018), 2655–2694.

[29] Heng Li, Weiyi Shang, Bram Adams, Mohammed Sayagh, and Ahmed E. Hassan. 2021. A Qualitative Study of the Benefits and Costs of Logging From Developers' Perspectives. *IEEE Trans. Software Eng.* 47, 12 (2021), 2858–2873. doi:10.1109/TSE.2020.2970422

[30] Heng Li, Weiyi Shang, Ying Zou, and Ahmed E. Hassan. 2017. Towards just-in-time suggestions for log changes. *Empirical Software Engineering* 22, 4 (2017), 1831–1865.

[31] Heng Li, Haoxiang Zhang, Shaowei Wang, and Ahmed E Hassan. 2021. Studying the practices of logging exception stack traces in open-source software projects. *IEEE Transactions on Software Engineering* 48, 12 (2021), 4907–4924.

[32] Shanshan Li, Xu Niu, Zhouyang Jia, Xiangke Liao, Ji Wang, and Tao Li. 2020. Guiding log revisions by learning from software evolution history. *Empir. Softw. Eng.* 25, 3 (2020), 2302–2340. doi:10.1007/S10664-019-09757-Y

[33] Shanshan Li, Xu Niu, Zhouyang Jia, Ji Wang, Haochen He, and Teng Wang. 2018. Logtracker: Learning log revision behaviors proactively from software evolution history. In *Proceedings of the 26th Conference on Program Comprehension*. 178–188.

[34] Yichen Li, Yintong Huo, Zhihan Jiang, Renyi Zhong, Pinjia He, Yuxin Su, Lionel C Briand, and Michael R Lyu. 2024. Exploring the effectiveness of LLMs in automated logging statement generation: An empirical study. *IEEE Transactions on Software Engineering* 50, 12 (2024), 3188–3207.

[35] Yichen Li, Yintong Huo, Renyi Zhong, Zhihan Jiang, Jinyang Liu, Junjie Huang, Jiazhen Gu, Pinjia He, and Michael R Lyu. 2024. Go Static: Contextualized Logging Statement Generation. ArXiv abs/2402.12958 (2024).

[36] Yichen Li, Jinyang Liu, Junsong Pu, Zhihan Jiang, Zhuangbin Chen, Xiao He, Tieying Zhang, Jianjun Chen, Yi Li, Rui Shi, et al. 2025. Automated Proactive Logging Quality Improvement for Large-Scale Codebases. In *2025 40th IEEE/ACM International Conference on Automated Software Engineering (ASE)*. IEEE, 3426–3437.

[37] Zhenhao Li, An Ran Chen, Xing Hu, Xin Xia, Tse-Hsun Chen, and Weiyi Shang. 2023. Are They All Good? Studying Practitioners' Expectations on the Readability of Log Messages. In *38th IEEE/ACM International Conference on Automated Software Engineering, ASE 2023, Luxembourg, September 11-15, 2023*. IEEE, 129–140. doi:10.1109/ASE56229.2023.00136

[38] Zhenhao Li, Tse-Hsun Chen, Jinqiu Yang, and Weiyi Shang. 2022. Studying Duplicate Logging Statements and Their Relationships With Code Clones. *IEEE Trans. Software Eng.* 48, 7 (2022), 2476–2494. doi:10.1109/TSE.2021.3060918

[39] Zhenhao Li, Tse-Hsun Chen, and Weiyi Shang. 2020. Where shall we log? studying and suggesting logging locations in code blocks. In *Proceedings of the 35th IEEE/ACM International Conference on Automated Software Engineering*.

361–372.

[40] Zhenhao Li, Tse-Hsun (Peter) Chen, Jinqiu Yang, and Weiyi Shang. 2019. Dlfinder: characterizing and detecting duplicate logging code smells. In *Proceedings of the 41st International Conference on Software Engineering, ICSE 2019, Montreal, QC, Canada, May 25-31, 2019*, Joanne M. Atlee, Tevfik Bultan, and Jon Whittle (Eds.). IEEE / ACM, 152–163. doi:10.1109/ICSE.2019.00032

[41] Zhenhao Li, Heng Li, Tse-Hsun Chen, and Weiyi Shang. 2021. Deeplv: Suggesting log levels using ordinal based neural networks. In *2021 IEEE/ACM 43rd International Conference on Software Engineering (ICSE)*. IEEE, 1461–1472.

[42] Yingwei Ma, Yongbin Li, Yihong Dong, Xue Jiang, Yanhao Li, Yue Liu, Rongyu Cao, Jue Chen, Fei Huang, and Binhua Li. 2025. Thinking longer, not larger: Enhancing software engineering agents via scaling test-time compute. In *2025 40th IEEE/ACM International Conference on Automated Software Engineering (ASE)*. IEEE, 3730–3741.

[43] Antonio Mastropaolo, Luca Pascarella, and Gabriele Bavota. 2022. Using deep learning to generate complete log statements. In *Proceedings of the 44th international conference on software engineering*. 2279–2290.

[44] Mary L. McHugh. 2012. Interrater reliability: the kappa statistic. *Biochemia Medica* 22, 3 (2012), 276–282.

[45] Microsoft. 2026. Architecture strategies for designing a monitoring system. https://learn.microsoft.com/en-us/azure/well-architected/operational-excellence/observability. Accessed: 2026-09-28.

[46] Microsoft. 2026. Avoid potentially expensive logging. https://learn.microsoft.com/en-us/dotnet/fundamentals/code-analysis/quality-rules/ca1873. Accessed: 2026-09-28.

[47] OpenAI. 2025. Codex CLI: A Lightweight Coding Agent That Runs in Your Terminal. https://github.com/openai/codex. Accessed: 2026-08-28.

[48] OpenAI. 2026. GPT-5.6. https://openai.com/index/gpt-5-6/.

[49] openhands. 2026. Openhands. https://github.com/OpenHands/OpenHands/releases.

[50] Youssef Esseddiq Ouatiti, Mohammed Sayagh, Hao Li, and Ahmed E Hassan. 2026. Do AI Coding Agents Log Like Humans? An Empirical Study. *arXiv preprint arXiv:2604.09409* (2026).

[51] Peter Pickerill, Heiko Joshua Jungen, Mirosław Ochodek, Michał Maćkowiak, and Miroslaw Staron. 2020. PHANTOM: Curating GitHub for engineered software projects using time-series clustering. *Empirical Software Engineering* 25, 4 (2020), 2897–2929.

[52] Qwen. 2026. Qwen3.8-Flash. https://modelstudio.console.alibabacloud.com/ap-southeast-1/model/market/detail/qwen3.8-flash?spm=a2c63.p38356.0.0.5b93776ffI0ynv&file=qwen3.8-flash&serviceSite=international. Last Access: 2026.

[53] Guoping Rong, Shenghui Gu, Haifeng Shen, He Zhang, and Hongyu Kuang. 2023. How do developers' profiles and experiences influence their logging practices? an empirical study of industrial practitioners. In *2023 IEEE/ACM 45th International Conference on Software Engineering (ICSE)*. IEEE, 855–867.

[54] Amazon Web Services. 2026. Logging best practices. https://docs.aws.amazon.com/prescriptive-guidance/latest/logging-monitoring-for-application-owners/logging-best-practices.html. Accessed: 2026-09-28.

[55] Weiyi Shang, Meiyappan Nagappan, and Ahmed E. Hassan. 2015. Studying the relationship between logging characteristics and the code quality of platform software. *Empir. Softw. Eng.* 20, 1 (2015), 1–27. doi:10.1007/S10664-013-9274-8

[56] SWE-agent Team. 2026. mini-SWE-Agent. https://mini-swe-agent.com/latest/.

[57] Boyin Tan, Junjielong Xu, Zhouruixing Zhu, and Pinjia He. 2025. Al-bench: A benchmark for automatic logging. *ACM Transactions on Software Engineering and Methodology* (2025).

[58] Yingchen Tian, Yuxia Zhang, Klaas-Jan Stol, Lin Jiang, and Hui Liu. 2022. What makes a good commit message?. In *Proceedings of the 44th International Conference on Software Engineering*. 2389–2401. doi:10.1145/3510003.3510205

[59] vLLM. 2026. vLLM. https://github.com/vllm-project. Accessed: 2026-09-28.

[60] Xingyao Wang, Boxuan Li, Yufan Song, Frank F Xu, Xiangru Tang, Mingchen Zhuge, Jiayi Pan, Yueqi Song, Bowen Li, Jaskirat Singh, et al. 2025. Openhands: An open platform for ai software developers as generalist agents, 2024. *URL https://arxiv. org/abs/2407.16741* 2, 5 (2025).

[61] Xin Wang, Zhenhao Li, and Zishuo Ding. 2025. Defects4Log: Benchmarking LLMs for logging code defect detection and reasoning. In *2025 40th IEEE/ACM International Conference on Automated Software Engineering (ASE)*. IEEE, 1931–1942.

[62] Fengcai Wen, Csaba Nagy, Michele Lanza, and Gabriele Bavota. 2022. Quick remedy commits and their impact on mining software repositories. *Empirical software engineering* 27, 1 (2022), 14.

[63] Xiaoyuan Xie, Zhipeng Cai, Songqiang Chen, and Jifeng Xuan. 2024. Fastlog: An end-to-end method to efficiently generate and insert logging statements. In *Proceedings of the 33rd ACM SIGSOFT International Symposium on Software Testing and Analysis*. 26–37.

[64] John Yang, Carlos Jimenez, Alexander Wettig, Kilian Lieret, Shunyu Yao, Karthik Narasimhan, and Ofir Press. 2024. Swe-agent: Agent-computer interfaces enable automated software engineering. *Advances in Neural Information Processing Systems* 37 (2024), 50528–50652.

[65] Ding Yuan, Soyeon Park, and Yuanyuan Zhou. 2012. Characterizing logging practices in open-source software. In *34th International Conference on Software Engineering, ICSE 2012, June 2-9, 2012, Zurich, Switzerland*, Martin Glinz, Gail C. Murphy, and Mauro Pezzè (Eds.). IEEE Computer Society, 102–112. doi:10.1109/ICSE.2012.6227202

[66] He Yang Yuan, Xin Wang, Kundi Yao, An Ran Chen, Zishuo Ding, and Zhenhao Li. 2026. Towards Secure Logging: Characterizing and Benchmarking Logging Code Security Issues with LLMs. *Proceedings of the ACM on Software Engineering* 3, FSE (2026), 745–765.

[67] Yi Zeng, Jinfu Chen, Weiyi Shang, and Tse-Hsun (Peter) Chen. 2019. Studying the characteristics of logging practices in mobile apps: a case study on F-Droid. *Empir. Softw. Eng.* 24, 6 (2019), 3394–3434. doi:10.1007/S10664-019-09687-9

[68] Haonan Zhang, Yiming Tang, Maxime Lamothe, Heng Li, and Weiyi Shang. 2022. Studying logging practice in test code. *Empirical Software Engineering* 27, 4 (2022), 83.

[69] Shaoqiu Zhang, Yuhang Wang, Jialiang Liang, Yuling Shi, Wenhao Zeng, Maoquan Wang, Shilin He, Ningyuan Xu, Siyu Ye, Kai Cai, et al. 2026. SWE-Explore: Benchmarking How Coding Agents Explore Repositories. *arXiv preprint arXiv:2606.07297* (2026).

[70] Xu Zhang, Yong Xu, Si Qin, Shilin He, Bo Qiao, Ze Li, Hongyu Zhang, Xukun Li, Yingnong Dang, Qingwei Lin, Murali Chintalapati, Saravanakumar Rajmohan, and Dongmei Zhang. 2021. Onion: identifying incident-indicating logs for cloud systems. In *ESEC/FSE '21: 29th ACM Joint European Software Engineering Conference and Symposium on the Foundations of Software Engineering, Athens, Greece, August 23-28, 2021*, Diomidis Spinellis, Georgios Gousios, Marsha Chechik, and Massimiliano Di Penta (Eds.). ACM, 1253–1263. doi:10.1145/3468264.3473919

[71] Chen Zhi, Shuiguang Deng, Junxiao Han, and Jianwei Yin. 2022. Towards automatic detection and prioritization of pre-logging overhead: a case study of hadoop ecosystem. *Automated Software Engineering* 29, 1 (2022), 11.

[72] Renyi Zhong, Yintong Huo, Wenwei Gu, Yichen Li, and Michael R Lyu. 2026. AutoLogger: A Multi-Agent Framework for the End-to-End Automated Logging. In *Proceedings of the 2026 34th IEEE/ACM International Conference on Program Comprehension*. 26–37.

[73] Renyi Zhong, Yichen Li, Jinxi Kuang, Wenwei Gu, Yintong Huo, and Michael R. Lyu. 2026. LogUpdater: Automated Detection and Repair of Specific Defects in Logging Statements. *ACM Trans. Softw. Eng. Methodol.* 35, 1 (2026), 16:1–16:31. doi:10.1145/3731754